\documentclass[12pt]{article}
\usepackage{cite}
\usepackage{amssymb}
\usepackage{pifont}\usepackage{amsmath}
\usepackage{subfigure}
\usepackage{graphics}
\usepackage{pstricks}
\usepackage{pst-plot}
\usepackage{pst-slpe}
\def\C{{\mathbf{C}}}
\def\eq#1{(\ref{#1})}

\def\pa{\partial}
\def\rt{\longrightarrow}
\def\R{{\mathbf{R}}}

\def\Z{{\mathbf{Z}}}
\begin{document}
\title{String networks as tropical curves}
\author{Koushik Ray \thanks{koushik@iacs.res.in}\\
\small Department of Theoretical Physics \&
\small Centre for Theoretical Sciences\\
\small Indian Association for
\small the Cultivation of Science\\
\small Calcutta 700 032, India.
}
\date{}
\maketitle
\begin{abstract}
\thispagestyle{empty}
\noindent 
A prescription for obtaining supergravity solutions for planar 
$(p,q)$-string networks is presented, based on earlier results. 
It shows that networks may be looked upon as 
tropical curves emerging as the spine of the amoeba of a holomorphic 
curve in M-theory. The
K\"ahler potential of supergravity
is identified with the corresponding Ronkin function. 
Implications of this identification in counting dyons is
discussed. 
\end{abstract}
String or brane networks are configurations in string theory made up of
intersecting strings or branes,
preserving lesser supersymmetry compared to a single brane.
Networks of electrically as well as magnetically charged
$(p,q)$-strings, in particular,  
have had a crucial role in the understanding of duality symmetries
of string theory since the early days of M- and F-theories
\cite{sch1,sen1,senF,km1,km2}. 
Following the identification of networks of
strings or branes as dyonic states in string theory, which in
turn are related by duality symmetries to black holes, the counting of
such networks have recently become important for a microscopic understanding of 
black hole entropy from string theory 
\cite{dvv,gaiotto,ssy1,ssy2, sen2,sen3}. 

String or brane networks may be studied within different frameworks,
namely, the world-sheet description of string theories, the
world-volume theories of D-branes or supergravity. String networks
have further been studied in M-theory \cite{kroghlee}. 

In this note we present a prescription for obtaining  a supergravity 
solution for general $(p,q)$-string
networks, directly related to their M-theoretic description as wrapped
membranes. The outline of the procedure is as follows. 
A $(p,q)$-string of type-IIB string theory
is described as a membrane in M-theory with
one circle wrapped on a torus. A network, in this setting, is
characterized by a holomorphic curve, 
called a spectral curve or a brane profile, written in suitable 
coordinates \cite{lunin,kroghlee}.
The solution proposed here hinges on the observation 
that a precise description of planar string networks 
may be given as a tropical curve \cite{gathman,mikhalkin1,mikhalkin2} 
corresponding to the spectral curve. 
The asymptotic $(p,q)$ charges that characterize a network are given
as the degree of the tropical curve \cite{mikhalkin3}.
The tropical curve is obtained as the spine of the amoeba
of the spectral curve \cite{rullgard1}.
The nexus between the M-theoretic and the tropical descriptions provides an
inkling of the choice of the K\"ahler potential in the
eleven-dimensional supergravity. 
We identify the Ronkin function of the amoeba
\cite{gathman,mikhalkin1,mikhalkin2,rullgard1} 
as the sole contribution to the K\"ahler potential due to the network,
thereby drastically simplifying the corresponding Monge-Amp\`ere
equation. This leads to an explicit solution for \emph{any} planar
$(p,q)$-string network and relates the membrane description of
networks with the supergravity description in eleven dimensions.

The identification of string networks as tropical curves
has interesting implications for their counting, which in
turn yields the degeneracy of certain 1/4-BPS dyons
\cite{dvv,gaiotto,ssy1,ssy2, sen2,sen3}. 
Tropical curves are
duals to subdivisions of Newton polygons \cite{gathman}. 
Thus the number of tropical curves of a given degree is related to the
number of regular subdivisions of the corresponding Newton polygon. 
While so-called singular tropical curves may be
conceived, corresponding to non-regular subdivisions, they will not
correspond to configurations with three-string junctions only. 
Henceforth a tropical curve will refer to a non-singular one. 
Thus, the degeneracy of $(p,q)$-string networks, for a given set of
asymptotic charges, receives a combinatorial description under the
above-mentioned identification.

We shall begin with a brief discussion of tropical curves
and their Ronkin function. We then identify
a planar string network in its membrane avatar 
as a tropical curve.  
Relating the K\"ahler potential of the eleven-dimensional supergravity
to the Ronkin function
of the spectral curve then leads to a very simple, but general, 
explicit solution. We close with a discussion
of applications to the counting problem.

Let us recall the description of the amoeba of a complex curve
and its spine \cite{gathman}. Let us consider a 
curve $\mathcal{C}$ in the affine space $\C^2$, with coordinates
$(u^1, u^2)$, given by a polynomial equation
\begin{equation}
\label{defc}
\mathcal{C} =
\left\{(u^1,u^2)|f(u^1,u^2)=\sum\limits_{i,j\in\mathbf{N}} a_{ij}
(u^1)^i(u^2)^j=0\right\},
\end{equation} 
where $a_{ij}$ are complex coefficients
and $\mathbf{N}$ denotes the set of
natural numbers. The curve $\mathcal{C}$ is first restricted
to $(\C^{\star})^2$, where $\C^{\star}$ denotes the complex plane sans 
the origin.
The restricted set is mapped, in turn, to the real plane by the Log-map,
\begin{equation} 
\label{logmap}
\begin{split}
\mathrm{Log} : (\C^{\star})^2 &\rt\R^2 \\
u=(u^1,u^2)\longmapsto(x^1,x^2) &:=(\log |u^1|, \log |u^2|).
\end{split}
\end{equation} 
The resulting subset $\mathcal{A}_{\mathcal{C}} =
\mathrm{Log}(\mathcal{C}\cap (\C^{\star})^2)$ of $\R^2$ is called the
\emph{amoeba} of the curve $\mathcal{C}$.
A family of amoebas parametrized by a small real number $\zeta$
is obtained by considering the Log-map with base
$\zeta$ as
\begin{equation} 
\begin{split}
\mathrm{Log}_{\zeta} : (\C^{\star})^2 &\rt\R^2 \\
u=(u^1,u^2)\longmapsto(x^1,x^2) &:=(-\log_{\zeta} |u^1|, -\log_{\zeta}
|u^2|).
\end{split}
\end{equation} 
The definition \eq{logmap} corresponds to $\zeta=1/e$. 
Reducing the family parameter results in shrinking the amoeba. In the
limit of vanishing $\zeta$ we obtain the spine of the amoeba
\cite{gathman,rullgard1}, called the
\emph{tropical curve}, denoted $\mathcal{C}_T$, corresponding to 
the curve $\mathcal{C}$. 

While this analytic definition is intuitively appealing, it is often
easier to enumerate tropical curves using a more combinatorially
amenable algebraic definition.  Given an algebraic curve
$\mathcal{C}$ as in \eq{defc} one first defines the function 
\footnote{Formally, 
the coefficients $a_{ij}$ appearing in \eq{defc}
are taken to be valued in a formal power-series,
the so-called Puiseux series
\cite{gathman}. However, this difference will be inconsequential for
our purposes here. }, 
\begin{equation}
\label{defct}
g(x^1,x^2) := \max\{ix^1+jx^2-\mathrm{val}(a_{ij}),
(i,j)\in\mathbf{N}^2, a_{ij}\neq 0\}, 
\end{equation} 
where $\mathrm{val}$, called the valuation, is an indicial
weight assigned to the coefficients. 
The tropical curve $\mathcal{C}_T$ is
the corner locus of this convex piecewise linear function, that is the
locus of points at which $g$ is not differentiable. 
Given a complex curve $\mathcal{C}$, both the definitions 
yield the same tropical curve.

\begin{figure}
\subfigure[Amoeba of the curve $u^1+u^2=1$]{%
\begin{pspicture}(-.5,0)(3,4)
\scalebox{.5}{\includegraphics{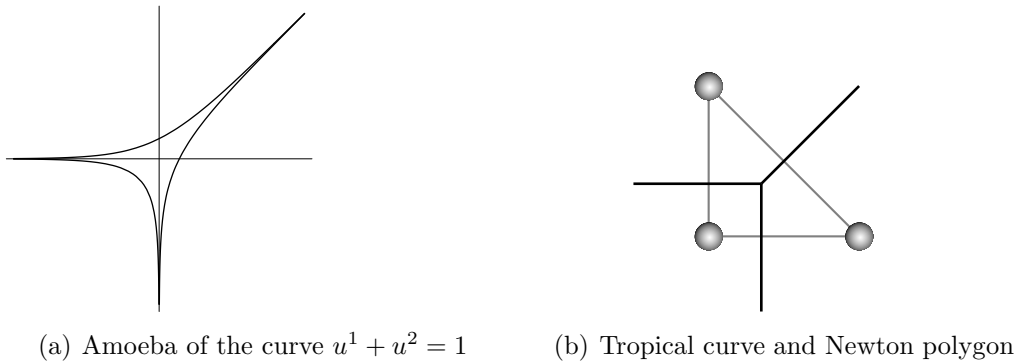}}
\end{pspicture}
\label{fig:amba}
}%
\subfigure[Tropical curve and Newton polygon]{
\label{fig:netwrk}
\begin{pspicture}(-2,-1)(4,4)
\psset{unit=2cm}
\psline[linecolor=gray](0,0)(0,1)
\psline[linecolor=gray](0,0)(1,0)
\psline[linecolor=gray](0,1)(1,0)
\rput(0,0){\pscircle[linestyle=none,fillstyle=ccslope,
slopeend=black,slopebegin=white,slopecenter=.54 .46](0,0){.1}}
\rput(1,0){\pscircle[linestyle=none,fillstyle=ccslope,
slopeend=black,slopebegin=white,slopecenter=.54 .46](0,0){.1}}
\rput(0,1){\pscircle[linestyle=none,fillstyle=ccslope,
slopeend=black,slopebegin=white,slopecenter=.54 .46](0,0){.1}}
\psline[linewidth=1pt](0.35,0.35)(0.35,-.5)
\psline[linewidth=1pt](0.35,0.35)(-.5,.35)
\psline[linewidth=1pt](0.35,0.35)(1,1)
\end{pspicture}
}
\caption{Amoeba and tropical curve}
\end{figure}

Let us illustrate the definitions with the simplest example
of the curve 
\begin{equation} 
\label{ex:crv}
\mathcal{C}=\{(u^1,u^2)\in\C^2|u^1+u^2=1\}. 
\end{equation} 
The amoeba of this curve obtained from 
\eq{logmap} is plotted in Figure~\ref{fig:amba}.
The tropical curve is
obtained from the combinatorial definition. The function $g$
becomes  
\begin{equation}
g(x^1,x^2)= \max (x^1,x^2,0),
\end{equation} 
yielding three line segments 
$\{x^1=0, x^2<0\}$, 
$\{x^1<0, x^2=0\}$, 
$\{x^1=x^2>0\}$. 
The corresponding tropical curve is the tree shown in
Figure~\ref{fig:netwrk}. 
It clearly is the limit as the amoeba in Figure~\ref{fig:amba}
shrinks to its spine. 
The between  the tree and a basic
three-string junction consisting of strings of charges $(0,-1)$,
$(-1,0)$ and $(1,1)$ is conspicuous from Figure~\ref{fig:netwrk}. 
Indeed, as we shall note below, planar string networks
constructed from $(p,q)$-strings can be defined as tropical
trees in general.

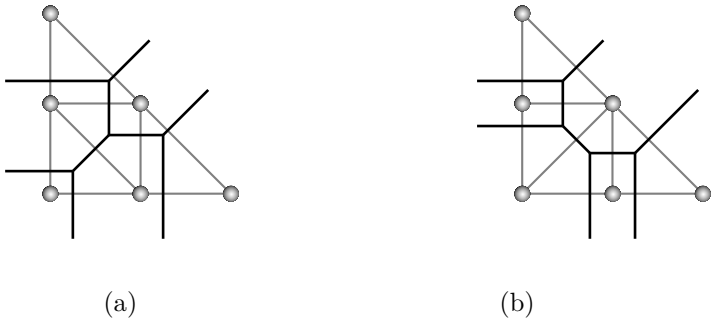
\begin{figure}[h]
\subfigure[]{%
\label{fig:tree1}
\begin{pspicture}(-2,-1)(4,4)
\psset{unit=1.2cm}
\psline[linecolor=gray](0,0)(0,2)
\psline[linecolor=gray](0,0)(2,0)
\psline[linecolor=gray](0,2)(2,0)
\psline[linecolor=gray](0,1)(1,1)
\psline[linecolor=gray](1,1)(1,0)
\psline[linecolor=gray](0,1)(1,0)
\rput(0,0){\pscircle[linestyle=none,fillstyle=ccslope,
slopeend=black,slopebegin=white,slopecenter=.54 .46](0,0){.1}}
\rput(1,0){\pscircle[linestyle=none,fillstyle=ccslope,
slopeend=black,slopebegin=white,slopecenter=.54 .46](0,0){.1}}
\rput(0,1){\pscircle[linestyle=none,fillstyle=ccslope,
slopeend=black,slopebegin=white,slopecenter=.54 .46](0,0){.1}}
\rput(2,0){\pscircle[linestyle=none,fillstyle=ccslope,
slopeend=black,slopebegin=white,slopecenter=.54 .46](0,0){.1}}
\rput(0,2){\pscircle[linestyle=none,fillstyle=ccslope,
slopeend=black,slopebegin=white,slopecenter=.54 .46](0,0){.1}}
\rput(1,1){\pscircle[linestyle=none,fillstyle=ccslope,
slopeend=black,slopebegin=white,slopecenter=.54 .46](0,0){.1}}
\psline[linewidth=1pt](0.25,0.25)(0.25,-.5)
\psline[linewidth=1pt](0.25,0.25)(-.5,.25)
\psline[linewidth=1pt](0.25,0.25)(.65,.65)
\psline[linewidth=1pt](.65,.65)(.65,1.25)
\psline[linewidth=1pt](.65,1.25)(-.5,1.25)
\psline[linewidth=1pt](.65,1.25)(1.1,1.7)
\psline[linewidth=1pt](.65,.65)(1.25,.65)
\psline[linewidth=1pt](1.25,.65)(1.25,-.5)
\psline[linewidth=1pt](1.25,.65)(1.75,1.15)
\end{pspicture}
}%
\subfigure[]{
\label{fig:tree2}
\begin{pspicture}(-2,-1)(2,4)
\psset{unit=1.2cm}
\psline[linecolor=gray](0,0)(0,2)
\psline[linecolor=gray](0,0)(2,0)
\psline[linecolor=gray](0,2)(2,0)
\psline[linecolor=gray](0,1)(1,1)
\psline[linecolor=gray](1,1)(1,0)
\psline[linecolor=gray](0,0)(1,1)
\rput(0,0){\pscircle[linestyle=none,fillstyle=ccslope,
slopeend=black,slopebegin=white,slopecenter=.54 .46](0,0){.1}}
\rput(1,0){\pscircle[linestyle=none,fillstyle=ccslope,
slopeend=black,slopebegin=white,slopecenter=.54 .46](0,0){.1}}
\rput(0,1){\pscircle[linestyle=none,fillstyle=ccslope,
slopeend=black,slopebegin=white,slopecenter=.54 .46](0,0){.1}}
\rput(2,0){\pscircle[linestyle=none,fillstyle=ccslope,
slopeend=black,slopebegin=white,slopecenter=.54 .46](0,0){.1}}
\rput(0,2){\pscircle[linestyle=none,fillstyle=ccslope,
slopeend=black,slopebegin=white,slopecenter=.54 .46](0,0){.1}}
\rput(1,1){\pscircle[linestyle=none,fillstyle=ccslope,
slopeend=black,slopebegin=white,slopecenter=.54 .46](0,0){.1}}
\psline[linewidth=1pt](0.75,0.45)(0.75,-.5)
\psline[linewidth=1pt](0.45,0.75)(-.5,.75)
\psline[linewidth=1pt](0.45,0.75)(.75,.45)
\psline[linewidth=1pt](.45,.75)(.45,1.25)
\psline[linewidth=1pt](.45,1.25)(-.5,1.25)
\psline[linewidth=1pt](.45,1.25)(.9,1.7)
\psline[linewidth=1pt](.75,.45)(1.25,.45)
\psline[linewidth=1pt](1.25,.45)(1.25,-.5)
\psline[linewidth=1pt](1.25,.45)(1.95,1.15)
\end{pspicture}
}
\caption{Tropical curves corresponding to different 
subdivisions of a Newton polygon}
\label{fig:quad}
\end{figure}

We can associate a Newton polygon to the curve $\mathcal{C}$ 
as the convex hull of the lattice points $(i,j)$ in $\R^2$ 
appearing in \eq{defc}. The tropical curve is obtained as the dual,
that is by drawing line segments in the plane perpendicular to the
lines in a regular subdivision of the Newton polygon.
For the above example, the Newton polygon has the lattice points
$(0,0)$, $(0,1)$ and $(1,0)$ as its vertices, as shown in
Figure~\ref{fig:netwrk}.
Different regular subdivisions of the Newton polygon
yields different tropical curves, with the same degree (number of
external legs) as illustrated in Figure~\ref{fig:quad}, where
tropical curves corresponding to the most general  quadratic plane
curve are shown.

Finally, let us recall that
the Ronkin function associated to the
curve $\mathcal{C}$, or, equivalently, to the
amoeba $\mathcal{A}_{\mathcal{C}}$, is defined as \cite{kos}
\begin{equation}
N_f(x^1,x^2)= \frac{1}{(2\pi i)^2} \int\limits_{|u^1|=e^{x^1}\atop
|u^2|=e^{x^2}} \log f(u_1,u_2)
\frac{du^1}{u^1}\frac{du^2}{u^2} 
\end{equation} 
The Ronkin function is convex over the amoeba and linear over each
component of  its complement \cite{mikhalkin2}.
The Ronkin function of a Harnack curve \cite{rullgard2}, 
which we assume of the curves that
we consider here, thereby guaranteeing maximality of the area of the
amoeba in the Lebesgue measure, satisfies the Monge-Amp\`ere equation
\cite{kos},
\begin{equation}
\label{monge}
\frac{\pa^2N_f}{\pa x^1\pa x^1} 
\frac{\pa^2N_f}{\pa x^2\pa x^2} 
-\frac{\pa^2N_f}{\pa x^1\pa x^2} 
\frac{\pa^2N_f}{\pa x^2\pa x^1} =\frac{1}{\pi^2}.
\end{equation} 

Let us now  go over to the description of string networks from
membranes in M-theory \cite{kroghlee}, envisaged as tropical curves.  
We shall consider the
eleven-dimensional supergravity limit of M-theory
on $\R^{1,8}\times T^2$, where the torus
$T^2$ is parametrized by the coordinates $x^3$ and $x^{10}$, with
periodicities\footnote{The convention for periodicities of the 
torus coordinates are chosen after \cite{lunin} and
differ from the one in \cite{kroghlee}. 
But this is only a change of coordinates,
resulting in a change of basis of the homology cycles of $T^2$, which
does not affect the physical properties. The string coupling is also
set to unity.} 
\begin{equation}
\label{period}
(x^3,x^{10})\sim(x^3+2\pi R,x^{10})\sim (x^3+ 2\pi R, x^{10}+2\pi R).
\end{equation} 
For a finite $R$ this describes type-IIB theory on a circle, while in the
limit of vanishing $R$ one recovers the type-IIB theory in ten
dimensions. A network of $(p,q)$-string lying in the 
$(x^1,x^2)$-plane is described in this setting through an auxiliary
curve, holomorphic, in order to be supersymmetric. We can define complex
coordinates
\begin{equation}
\label{zcoor}
z^1 = x^1+ix^3,\quad z^2=x^2+ix^{10}.
\end{equation} 
parametrizing $\C^2$. Then the coordinates 
\begin{equation}
u^1 = e^{z^1/R} ,\quad u^2=e^{-z^2/R}.
\end{equation} 
parametrize $(\C^{\star})^2$. 
A single $(p,q)$-string, lying in the $(x^1,x^2)$-plane is 
specified by a holomorphic curve in the $(u^1,u^2)$ coordinates as
\begin{equation}
(u^1)^p(u^2)^q=1.
\end{equation} 
The string itself may be obtained from this as the tropical limit
\begin{equation}
px^1=qx^2,
\end{equation}
following the algebraic definition. 
More generally, a string network is specified by giving a spectral
curve \cite{kroghlee}
\begin{equation}
f(u^1,u^2)=\sum_{p,q\in\Z} (u^1)^p (u^2)^q =0.
\end{equation} 
The spectral curve describes a membrane that wraps on the torus
$T^2$ yielding the string network  in the tropical
limit. The description of networks as tropical curves coincides with
the more traditional picture obtained by analyzing the asymptotics
\cite{kroghlee}. 

Let us now consider the supergravity
description of string networks. Different aspects of supergravity
configurations corresponding to networks of strings and branes
have been extensively studied
\cite[and reference therein]{lunin}. 
Here we only quote a few formulas relevant for the
case at hand, namely, the M-theoretic description of planar string
networks at low energies.
We seek  a  metric
as well as associated fluxes corresponding to a string network in the
eleven-dimensional supergravity. Solutions in type-IIB string theory
ensues in the limit of vanishing  $R$.
upon assuming isometries along two directions $x^3$ and $x^{10}$.
The metric ans\"atz for the eleven-dimensional geometry,
corresponding to an M-theory configuration 
with $U(1)_t\times SO(6)$ symmetry and 
preserving eight supercharges is 
\cite{lunin}
\begin{gather}
ds^2 = -e^{2A}dt^2+2e^{2A}h_{a\bar{b}}dz^a d\bar{z}^b +
e^{-A}(dy^2+y^2d\Omega_5^2),\\
h_{a\bar{b}} = \frac{\pa^2 K}{\pa z^a\pa {\bar{z}^b}},
\end{gather}
where $K$ denotes the K\"ahler potential. The string lies within a
four-dimensional subspace of the eleven-dimensional space-time, 
with complex coordinates as in \eq{zcoor}, where
$x^{10}$ denotes the coordinate of the eleventh dimension of
M-theory. Also, $y$  and $\Omega_5$ denote, respectively,
the radial and angular coordinates of
the six-dimensional part of the space-time transverse to $\C^2(z^1,z^2)$,
while $t$ denotes the temporal coordinate.
The fluxes are determined in terms of  the function $A$
appearing in the metric and the K\"ahler potential. 
In order for a network lying in the $x^1$-$x^2$-plane to
preserve some supersymmetry, the K\"ahler potential $K$ is required to
satisfy \cite{lunin}
\begin{gather}\label{mamp}
\frac{\pa^2 K}{\pa z^1\pa {\bar{z}}^1}
\frac{\pa^2 K}{\pa z^2\pa {\bar{z}}^2}
-\frac{\pa^2 K}{\pa z^1\pa {\bar{z}}^2}
\frac{\pa^2 K}{\pa z^2\pa {\bar{z}}^1} = \frac{1}{4} e^{-3A},\\
\label{mampy}
\nabla_y K = -2 e^{-3A},
\end{gather}
where $\nabla_y$ denotes the $y$-Laplacian.

Assuming a further $U(1)^2$ isometry, corresponding to
periodicities along the coordinates $x^3$ and $x^{10}$ of
the torus $T^2$, \eq{mamp} reduces to
\begin{equation}
\label{mampx}
\frac{\pa^2 K}{\pa x^1\pa x^1} 
\frac{\pa^2 K}{\pa x^2 \pa x^2} 
-\frac{\pa^2 K}{\pa x^1\pa x^2} 
\frac{\pa^2 K}{\pa x^2\pa x^1} 
= \frac{1}{4} e^{-3A}.
\end{equation} 
Given a particular network, a solution to this determines its effect on
the geometry of the target space. 
A simple solution arises by 
comparing \eq{mampx} and  \eq{monge}. 
We write the K\"ahler potential using the Ronkin
function corresponding to the spectral curve $\mathcal{C}$ as 
\begin{equation}
K=N_f(x^1/R,-x^2/R) + \psi(y),
\end{equation} 
where now $R$ is a function of $y$ only and so is $\psi$, 
thanks to the $SO(6)$ symmetry. 
Moreover, $R(y)$ is stipulated to vanish at a single value of $y$
signalling the presence of a source. 
Using the expressions $|u^1| = e^{x^1/R}$ and
$|u^2|=e^{-x^2/R}$, we note that the K\"ahler potential satisfies, 
\begin{equation}
\frac{\pa^2 K}{\pa x^1\pa x^1} 
\frac{\pa^2 K}{\pa x^2\pa x^2} 
-\frac{\pa^2 K}{\pa x^1\pa x^2} 
\frac{\pa^2 K}{\pa x^2\pa x^1} 
= \frac{1}{\pi^2 R^4},
\end{equation} 
in view of \eq{monge}.
This, in turn, constrains the function $A$ to be a function of $y$
alone, by
\eq{mampy}, such that
$e^{-3A}=4/\pi^2R^4$ while the function $\psi(y)$ now 
satisfies the six-dimensional Laplace equation
\begin{equation}
\label{eq:psi}
\nabla_yK = -\frac{8}{\pi^2R^4}.
\end{equation} 

Intuitively, in the presence of network sources 
the geometry is expected to be shaped after the amoeba.
The identification of string networks as tropical curves 
and writing the K\"ahler potential in terms of the Ronkin function 
realizes this, yielding a
simple solution to \emph{all} planar string networks. The
K\"ahler potential and hence the metric $h_{a\bar b}$ depends on the
specific network chosen as the Ronkin function is
associated to the spectral curve. The network affects the transverse 
part of the space-time through $R$. 
The $y$-dependence of the K\"ahler function is
not fixed at this stage, due to the provision of adding an
arbitrary function of $y$ to the K\"ahler potential \cite{lunin}.
Fixing it requires imposing specific boundary conditions. 
The amoeba of the curve $\mathcal{C}$ goes over to the
string network in the limit of vanishing $R$, if we identify the 
parameter $\zeta$ as $\zeta = e^{-1/R}$. 

Let us now briefly indicate some consequences of these considerations.
As illustrated in Figure~\ref{fig:quad}, string networks with
specified asymptotic charges, looked upon as tropical curves 
with a specified degree corresponding to external legs, 
may differ in the internal structure. Two different networks may
thus correspond to the same spectral curve as the spectral curve 
only determines the
Newton polygon, and not its triangulation. A Newton polygon
admits, more often than not, various subdivisions. Hence more than
one tropical curves, all with the same degree (external legs), 
correspond to the 
same Newton polygon by duality, mentioned earlier. Translated to
networks, this implies that there is a degeneracy of string networks
with specified asymptotic charges, corresponding to a spectral curve. 
With our identification of networks as tropical curves, now, the
degeneracy equals 
the number of regular subdivisions of the Newton polygon corresponding
to the spectral curve. While there seems to be no general formula for
the number of subdivisions of polygons, some estimates exist
\cite{ziegler,mikhalkin4}, especially for polygons of small size, as
well as other numerical means, which are now at our disposal.
As a simple application let us note that an $SL(2,Z)$
transformation does not change the volume of the Newton polygon of
Figure~\ref{fig:netwrk}. Hence under this transformation the 
Newton pollygon does not
pick up any extra lattice point. It follows that the number of
subdivisions does not change and we conclude that
an S-duality
transformation of the basic three-string junction shown in
Figure~\ref{fig:netwrk} does not alter the degeneracy \cite{sen3}.
On the other hand, the generic spectral curve corresponding to the
configurations in Figure~\ref{fig:quad} is a quadric. 
This network does not belong to the S-duality orbit of the basic
junction of Figure~\ref{fig:netwrk} \cite{sen3}. Hence the known
string theoretic formulas can not be used to calculate its degeneracy.
However, the number of regular
subdivisions of the corresponding Newton polygon 
is two, predicting a doubly degenerate string network. 

To conclude, in this note we have presented a simple prescription for
obtaining  supergravity solutions of planar $(p,q)$-string networks.
It incorporates the M-theoretic membrane description of
the networks in terms of spectral curves.
A network is defined as the spine of the amoeba 
of the spectral curve. The verisimilitude of string networks and grid
diagrams with tropical curves have been noted earlier
\cite{nekra,eto,hanany,aharony,kol,mikhalkin3}. However, obtaining a 
supergravity solution
ensuing from identifying them establishes a precise connection. 
The prescription works for all planar networks.
Furthermore, the networks are ``sensed" by the transverse
coordinates solely by the presence of a source as a
singularity in \eq{eq:psi}. The solution is generic in this sense. 
Defining networks as tropical curves have important consequences
in estimating the degeneracy of networks combinatorially. 
We hope that this will be useful in studying the 
degeneracies in general. 
Generalizations to higher dimensional networks
\cite{km1,km2,lunin} appears possible and will be reported separately. 
\section*{Acknowledgement}
It is a pleasure to thank Alok Kumar and Subir Mukhopadhyay for useful
discussions and  careful reading of the manuscript.

\end{document}